\documentclass[trackchanges]{aastex701}

\usepackage{comment}

\begin{document}

\title{An Atlas of Spectroheliograms from 3641 to 6600~\AA}

\author[]{P. Váradi Nagy}
\affiliation{Unaffiliated, Cluj-Napoca, Romania}
\email[show]{palnagy@gmail.com}  

\author[orcid=0000-0002-0484-7634]{A.G.M. Pietrow} 
\affiliation{Leibniz-Institut für Astrophysik Potsdam (AIP), An der Sternwarte 16, 14482 Potsdam, Germany}
\email{apietrow@aip.de}

\begin{abstract}

We present a spectral atlas of solar spectroheliograms covering the wavelength range from 3641 to 6600~\AA, with continuous coverage between 3711 and 5300~\AA, and sparser coverage beyond this range. 
The spectral resolution varies between R $\sim$ 20 000 and 40 000, with a spectral step size between 60 and 90~m\AA, while the spatial resolution averages around 2.5 arcseconds. 
These observations were acquired over three months during the 2025 solar maximum, using amateur spectroheliographs (Sol'Ex and ML Astro SHG 700). The atlas is accessible via an interactive online platform with navigation tools and direct access to individual spectroheliograms. 
\end{abstract}


\section{Introduction} 

There is renewed interest in the broad solar spectrum, driven by the newest generation of nighttime telescopes, which are now not only capable of detecting stellar activity but are also limited by it \citep[e.g.][]{Radica2025}. This limitation, in combination with the advent of so-called Sun-as-a-star telescopes that take disk-integrated solar spectra over wide wavelength ranges \citep[e.g.][]{2023Zhao}, has spearheaded comparative studies between solar and stellar observations. This is because the brightness and relative size of the Sun allow for detailed studies into its disk-integrated behavior. While detailed spectral atlases \citep[e.g.]{Kurucz1984, Ellwarth2023} and line lists \citep[e.g.][]{Moore1966} exist, so far only limited collections of spectroheliograms were published \citep[e.g.][]{Title1966}. Our new `Solar Spectroheliogram Atlas' fills this gap by providing a continuous set of full-disk spectroheliograms between 3711 and 5300~\AA, selected spectral regions between 5300 and 6600~\AA, and additional coverage of the bluer wavelengths around the Balmer jump (3650~\AA).

\begin{figure*}[ht!]
\centering
\includegraphics[width=0.9\columnwidth, trim={0 0 0 0},clip]{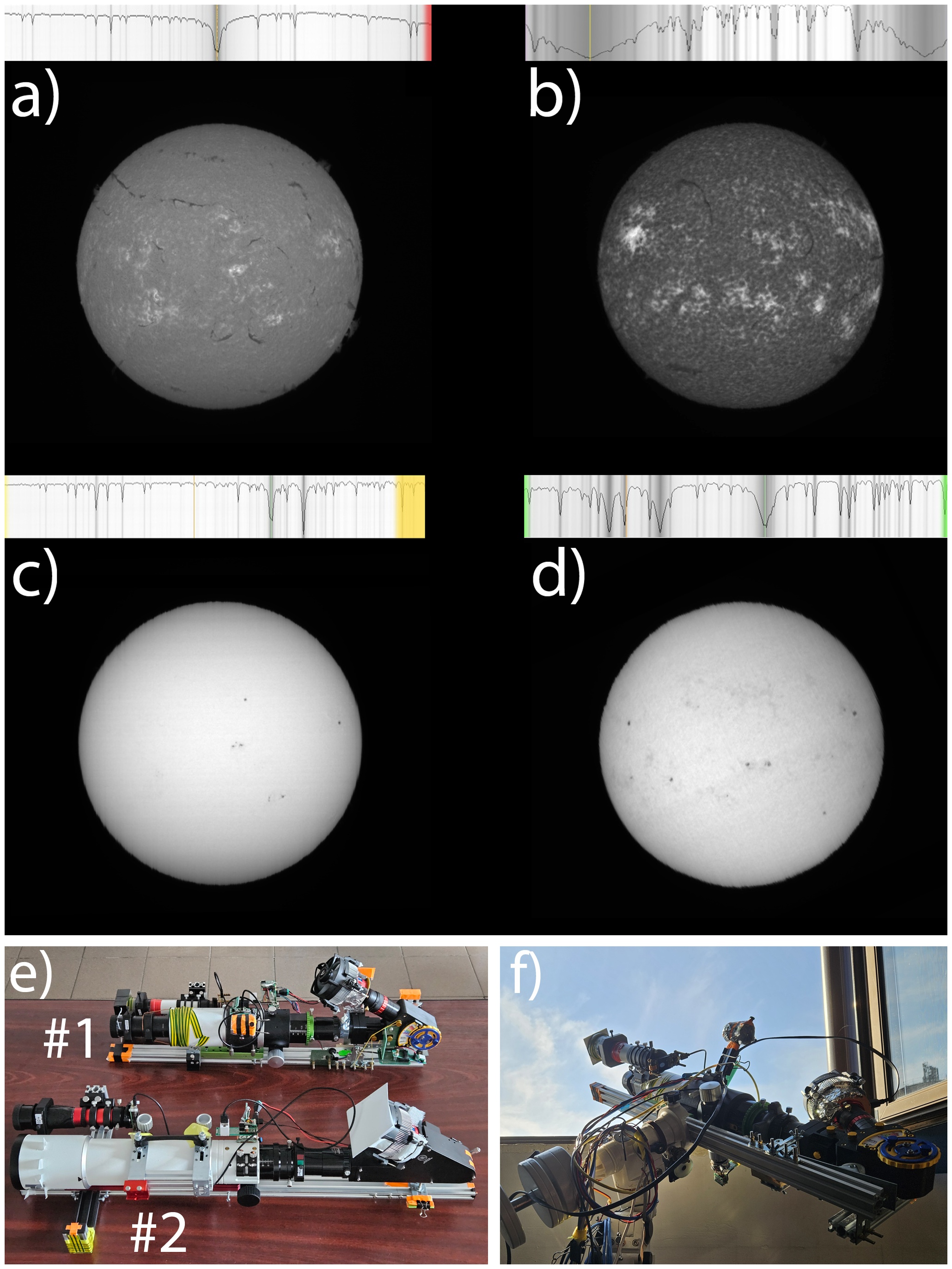}
\caption{A set of four spectroheliograms with corresponding cube spectra from the atlas, and the instruments used to capture them. The respective spectroheliograms correspond to the location of the orange line on the spectrum. The solar disks are tuned to a) H$\alpha$ 6563~\AA, b) Ca II K 3934~\AA, c) He D3 5876~\AA, and d) Fe II 5169~\AA.  The instruments shown in panel e) are the Sol'Ex ($\#$1) and the SHG 700 ($\#$2), with panel f) showing the Sol'Ex mounted and scanning the solar disk. \label{fig:general}}

\end{figure*}

\section{Atlas details}
The atlas consists of approximately 50 000 spectroheliograms, comprising $\sim$4.5 GB of compressed data, taken between 3641 and 6600~\AA. The spectral resolution is not consistent over the wavelength range, resulting in a step size of 60 to 90~m\AA\ throughout the atlas. The spatial resolution is often seeing limited at around 2.5 arcsec (See Fig. \ref{fig:general}a-d).  The coverage is continuous from 3711 to 5300~\AA\, with a sparser sampling outside this range. This is primarily due to the decreasing number of spectral lines and the increasing number of atmospheric contamination bands. Each spectral raster, or cube,  is between 20 to 70~\AA\ wide, with overlaps, and adjacent rasters are usually not taken on the same day. 

On the website\footnote{\url{https://csillagtura.ro/projektek/aladin-photo/solar-disks-spectrum-atlas}} the user is greeted with a broadband spectrum showing the current tuning. Below that is a set of commonly used wavelengths for quick navigation. Next is a set of spectral cubes that allow for more general wavelength selection. Once a cube is selected, 2D and 1D spectra are displayed. An orange marker denotes the selected wavelength and a green marker denotes the reference line of the disks' geometric reconstruction. Finally, the selected spectroheliogram is shown at the bottom. 
Users should be aware that data quality varies with atmospheric conditions and instrument stability. However, we consider the atlas scientifically and educationally valuable because of its novelty and scale. The data is also available for download on Zenodo \citep{varadi2025}.

\section{Observations and methods}

The data was taken between 2025-04-20 and 2025-07-06 in Cluj-Napoca, Romania. The site was chosen primarily for convenience and availability rather than for its astronomical properties. Observations were made when weather permitted and solar activity was present. The exact date is provided in the title of each image.

The observations were made with amateur instruments which were composed of a combination of off-the-shelf and custom-built parts. Both instruments are spectroheliographs (SHG) attached to common refractors (See Fig. \ref{fig:general}e,f), capable of observing the full solar disk in roughly one minute. The first instrument is a \textit{Solar Explorer}\footnote{\url{https://solex.astrosurf.com/}} \citep[or \textit{Sol'Ex}, ][]{2023Buil} on a 62/400 refractor, with the aperture limited to $\sim$4 cm by the energy rejection filters (ERF) in front of the objective (See Fig. \ref{fig:general}f). The second one is an 80/540 refractor with a $\sim$8cm clear aperture and a similar but more robust \textit{ML Astro SHG 700}\footnote{\url{https://mlastro.com/mlastro-shg}}. Both instruments use a grating of 2400 lines per mm, and can observe different wavelengths by adjusting the angle in their elbow structure. The \textit{Sol'Ex} and \textit{ML Astro} have a respective spectral resolution of R $\sim$ 40 000 and 20 000 at 6563~\AA.

The spectra were captured on a \textit{ZWO ASI 678MM}camera. In both cases, the optics were aligned so that the slit-grating-sensor ensemble was approximately perpendicular to the scanning direction, which was chosen to be the right ascension. Maintaining alignment and collimation is a challenge for these instruments, especially for the plastic of the \textit{Sol'Ex}, which needed an additional support structure built around it to limit sagging and keep its orientation, but to a smaller degree, even the focuser of the telescope is prone to produce image shift, due to the gears in the focusing mechanism. 

In order to reduce the thermal load on the slit, onto which the Sun is focused, both optical trains use bandpass filters and etalons as ERFs, these were placed in front of the refractor, limiting its aperture for the Sol'Ex; and in front of the slit but well out of focus for the SHG 700. 

A \textit{SkyWatcher EQ3} equatorial mount with a custom motor driver and firmware was used to automate the scanning routine and reduce vibrations during observations, achieving a raster cadence of around 1 minute. 

The entire optical assembly moves during the recording process, starting with a dark sky away from the Sun, after which it scans over the disk. This is similar to the way that the Chinese H$\alpha$ Solar Explorer satellite \citep{Chase2022} operates. The data was recorded using {\tt{SharpCap}} and saved in \textit{SER} format. Depending on the chosen wavelength range, several hundred full-disk spectroheliograms can be reconstructed from one take. The wavelength range was chosen as a compromise between the rolling shutter, camera scan lines count and readout speed, disk write speed, mount movement (i.e. angular speed of the scanning), and general sampling rate dictated by the exposure time while keeping the gain as low as possible.

The disks were reconstructed using the {\tt JSol'Ex} open source software\footnote{\url{https://melix.github.io/astro4j/latest/en/jsolex.html}}, which analyzes the recorded \textit{SER} file. It then finds the polynomial to straighten the bending (or smile) of the spectral lines (this is where the green marker comes from), and applies geometric corrections to turn the oval shaped scan into a round and centered solar disk that is rotated to be aligned by the solar equator. We explicitly avoided applying any cosmetic corrections (such as anti-jagging) and/or contrast enhancements (such as CLAHE). 

To extract a spectroheliogram for each wavelength, this process was automated with {\tt ImageMath}, the internal processing engine of {\tt JSol'Ex}. At the time of recording, there was no means to loop within this engine, therefore an external wrapper of scripts was used for this purpose. 

The wavelength calibration was done using the line list by  \citet{Moore1966} and the BASS2000\footnote{\url{https://bass2000.obspm.fr/solar_spect.php}} atlas, which is a derivative of the Liège atlas \citep{Delbouille1973}. 

Raw observations consist of multiple scans, ie video recordings for the same wavelength range, each recording about 10-15~GB. The best is selected for processing, resulting in about 20~GB of reconstructed fits files. To keep storage and presentation manageable, the resulting spectroheliograms and averaged spectra were downsampled to jpegs with a total compressed atlas size of around 4.5~GB. 
\clearpage
\begin{acknowledgments}
AP is supported by the \emph{Deut\-sche For\-schungs\-ge\-mein\-schaft, DFG\/} project number PI 2102/1-1

This work has made use of the VALD database, operated at Uppsala University, the Institute of Astronomy RAS in Moscow, and the University of Vienna.
\end{acknowledgments}

\bibliography{sample701}{}
\bibliographystyle{aasjournalv7}

\end{document}